\pdfoutput=1
\documentclass[letterpaper,titlepage,11pt]{article}

\usepackage{jheppub} 
\usepackage[T1]{fontenc}
\usepackage[utf8]{inputenc}
\usepackage[turkish, english]{babel} 

\usepackage[active]{srcltx}
\usepackage{textcomp}
\usepackage{amsmath}
\usepackage{amssymb}
\usepackage{esint}
\usepackage{multirow}
\usepackage{empheq}
\usepackage{textcomp}
\usepackage{xcolor}

\usepackage{enumerate}
\usepackage{xcolor}
\def\be{\begin{eqnarray}}
\def\ee{\end{eqnarray}}
\def\beann{\begin{eqnarray*}}
\def\eeann{\end{eqnarray*}}
\def\beq{\begin{equation}}
\def\eeq{\end{equation}}
\def\ba{\begin{array}}
\def\ea{\end{array}}
\def\ben{\begin{enumerate}}
\def\een{\end{enumerate}}
\def\bea{\begin{eqnarray}}
\def\eea{\end{eqnarray}}

\providecommand{\Lt}{{\tt L}}
\renewcommand{\Lt}{{\tt L}}
\providecommand{\Wt}{{\tt W}}
\renewcommand{\Wt}{{\tt W}}

\providecommand{\Tt}{{\tt T}}
\renewcommand{\Tt}{{\tt T}}

\def\be{\begin{equation}}
\def\ee{\end{equation}}
\def\bea{\begin{eqnarray}}
\def\eea{\end{eqnarray}}
\def\ba{\begin{array}}
\def\ea{\end{array}}

\def\tr{\text{tr}}

\usepackage{amsfonts}

\usepackage{tikz}

\usepackage{bbm}

\usepackage[dvipsnames]{xcolor}

\newcommand{\DOI}[1]{%
\href{https://doi.org/#1}{\textcolor{MidnightBlue}{doi:#1}}}

\newcommand{\ARXIV}[2]{%
\href{https://arxiv.org/abs/#1}{\textcolor{BrickRed}{arXiv:#1 [#2]}}}

\title{\boldmath\LARGE{{Generalized Schwarzian Dynamics from a Bulk-First BF Perspective}}}

\subheader{\today}
\newcommand{\itu}{\dagger}

\author[\itu]{H. T. \"Ozer}
\emailAdd{ozert@itu.edu.tr}
\author[\itu]{,\,\,\,Ayt\"ul Filiz}
\emailAdd{aytulfiliz@itu.edu.tr}

\affiliation[\itu]{Istanbul Technical University,\,Faculty of Science and Letters,
\,Physics Department,\\34469 Maslak,\,Istanbul,Turkey.}

\abstract{
We investigate the emergence of generalized Schwarzian dynamics from a bulk-first BF perspective. Starting from two-dimensional BF gravity, we analyze the associated boundary phase space and its Drinfeld--Sokolov reductions. For the $\mathfrak{sl}(2,\mathbb{R})$ theory, we recover the ordinary Schwarzian action as the reduced boundary dynamics arising from a particular sector of the BF asymptotic phase space. We then extend this construction to $\mathfrak{sl}(3,\mathbb{R})$, where the reduced dynamics is governed by the second and third Wilczynski invariants, providing a natural higher-rank generalization of the Schwarzian derivative. In this framework, generalized Schwarzian dynamics emerges directly from flat BF connections and their companion forms rather than being introduced as an independent boundary theory. We further relate the resulting projective invariants to Casimir charges, monodromy data, and generalized Schwarzian thermodynamics, including monodromy spectra and semiclassical thermodynamics. In particular, constant projective invariants determine the corresponding Casimir sectors and monodromy data, which in turn organize the thermodynamic structure of the theory. Our results provide a unified bulk-first description of Schwarzian and generalized Schwarzian dynamics and reveal a direct link between BF gravity, asymptotic symmetry reductions, projective geometry, and boundary thermodynamics.
}
\begin{document}
\maketitle
\flushbottom
\section{Introduction}
The Schwarzian theory has emerged as a central framework in the study of two-dimensional gravity, low-dimensional holography, and quantum chaotic systems. In the context of Jackiw--Teitelboim (JT) gravity, it provides an effective description of the low-energy boundary dynamics associated with nearly AdS$_2$ spacetimes and plays a prominent role in the connection between gravitational systems and the Sachdev--Ye--Kitaev (SYK) model~\cite{Almheiri:2014cka,Maldacena:2016upp,Engelsoy:2016xyb,Sachdev:2010um,Sachdev:2019bjn}. In its standard formulation, the Schwarzian action governs the dynamics of boundary reparametrizations and captures the leading effects of the spontaneous and explicit breaking of conformal symmetry. Owing to its remarkable simplicity and universality, the Schwarzian theory has become one of the most extensively studied effective descriptions of near-AdS$_2$ gravitational dynamics~\cite{Stanford:2017thb,Mertens:2017mtv,Cotler:2016fpe,Jensen:2016pah}.

While the Schwarzian action is commonly introduced as an effective boundary theory, several complementary approaches suggest that it may arise from a more fundamental gauge-theoretic description of two-dimensional gravity. In particular, Jackiw--Teitelboim gravity admits a BF formulation in which the fundamental fields consist of a flat gauge connection and an adjoint-valued dilaton field~\cite{Witten:1991zz,Ikeda:1993fh,Schaller:1994es}. Within this framework, the bulk dynamics is governed by flatness and covariant constancy conditions, while boundary observables emerge only after appropriate boundary conditions and phase-space reductions are imposed. From this viewpoint, the Schwarzian theory appears not as a primary ingredient of the gravitational system but rather as an effective description associated with a particular reduced sector of the underlying BF phase space.

This observation motivates a bulk-first perspective on boundary dynamics. Rather than beginning with an effective boundary action and subsequently identifying its bulk interpretation, one may instead start from the gauge-theoretic structure of BF gravity and investigate how boundary theories emerge from the associated asymptotic phase space. In such an approach, flat connections, residual gauge transformations, and boundary symmetry reductions constitute the primary ingredients of the construction, while effective boundary actions arise only after suitable reductions have been performed. This viewpoint shifts the emphasis from boundary reparametrization dynamics to the underlying gauge structure from which such dynamics originates~\cite{Gonzalez:2018enk,Ozer:2025bpb}. 
\color{black}
Early bulk-first derivations of Schwarzian dynamics from two-dimensional BF theory were developed in Refs.~\cite{Mertens:2018fds,Blommaert:2018oro}, where suitable BF boundary terms give rise to particle-on-a-group descriptions whose Hamiltonian reduction yields the Schwarzian theory. Higher-rank extensions of this perspective, including generalized Schwarzian dynamics associated with higher-spin BF theories, were subsequently explored in Ref.~\cite{Gonzalez:2018enk}.
\color{black}

While the emergence of the ordinary Schwarzian theory from BF gravity and related gauge-theoretic constructions is by now reasonably well understood, considerably less is known about analogous bulk-first descriptions of generalized Schwarzian dynamics~\cite{Maldacena:2016upp,Engelsoy:2016xyb,Gonzalez:2018enk}. In particular, higher-rank gauge algebras naturally give rise to enlarged asymptotic symmetry structures and additional boundary degrees of freedom, suggesting the existence of boundary theories extending the ordinary Schwarzian framework~\cite{Campoleoni:2010zq,Henneaux:2010xg,Gutperle:2011kf,Perez:2013xi,deBoer:2013gz}. However, a systematic derivation of such generalized Schwarzian dynamics directly from BF gravity, together with its geometric and thermodynamic interpretation, remains comparatively unexplored.

In this work, we develop a bulk-first construction of generalized Schwarzian dynamics within the framework of BF gravity. Starting from the $\mathfrak{sl}(3,\mathbb{R})$ BF theory~\cite{Campoleoni:2010zq,Henneaux:2010xg},
we investigate the corresponding asymptotic phase space and its Drinfeld--Sokolov reductions. We show that the resulting reduced dynamics is naturally governed by the second and third Wilczynski invariants, which generalize the role played by the ordinary Schwarzian derivative in the $\mathfrak{sl}(2,\mathbb{R})$ case. From this perspective, generalized Schwarzian dynamics emerges directly from flat BF connections and their reduced boundary structures rather than being introduced as an independent boundary theory.

Beyond the construction itself, our analysis establishes direct links between generalized Schwarzian dynamics, Casimir structures, and global boundary data. In particular, the Wilczynski invariants arising from the reduced BF description admit a natural interpretation in terms of quadratic and cubic Casimir sectors, allowing the generalized Schwarzian theory to be connected with monodromy data and associated thermodynamic observables. This framework provides a unified setting in which geometric invariants, asymptotic symmetry reductions, and boundary thermodynamics can be studied within a common bulk-based description~ \cite{OvsienkoTabachnikov,Wilczynski}.

This paper is organized as follows.
Section~\ref{sec:bf} reviews the BF formulation of two-dimensional
gravity and the role of boundary terms, flat connections, and residual
gauge symmetries.
Section~\ref{sec:schwarzian} derives the ordinary Schwarzian theory
from BF gravity through Drinfeld--Sokolov boundary conditions.
Section~\ref{sec:sl3} develops the
$\mathfrak{sl}(3,\mathbb{R})$ extension and its associated
$W_3$ structure.
Section~\ref{sec:wilczynski} derives the second and third Wilczynski
invariants from flat BF connections and their companion form.
Section~\ref{sec:gsch} constructs the corresponding generalized
Schwarzian actions in terms of the quadratic and cubic Casimir
sectors.
Section~\ref{sec:dilaton_stabilizers} discusses the dilaton sector and
its associated stabilizer structures in the spin-2 and spin-3
theories.
Section~\ref{sec:monodromy} analyzes the relation between Wilczynski
invariants, monodromy sectors, and boundary data.
Section~\ref{sec:entropy} studies the thermodynamic interpretation of
the generalized Schwarzian theory, including partition functions,
monodromy spectra, and semiclassical entropy.
Finally, Section~\ref{sec:discussion} summarizes the main results and
discusses possible extensions to higher-rank BF theories.

\section{BF Gravity and Boundary Dynamics}
\label{sec:bf}

BF theory provides a natural gauge-theoretic framework for describing two-dimensional gravity. In this formulation, the fundamental variables are a Lie-algebra-valued connection and an adjoint-valued scalar field, which together encode the geometric content of the theory. Rather than treating the metric as a primary variable, the BF description emphasizes the underlying gauge structure and formulates gravity as a constrained topological field theory. The resulting equations of motion impose flatness of the gauge connection and covariant constancy of the adjoint field, implying the absence of local propagating bulk degrees of freedom. Consequently, the physically relevant information is encoded in global and boundary data, making BF gravity particularly suitable for the study of asymptotic symmetries, boundary reductions, and emergent boundary dynamics~\cite{Witten:1991zz,Ikeda:1993fh,Schaller:1994es,Gonzalez:2018enk}.


The dynamics of BF gravity is encoded in a remarkably simple action constructed from the pairing between the adjoint field and the curvature of the gauge connection. For a gauge algebra $\mathfrak{g}$ equipped with a non-degenerate invariant bilinear form. 
Jackiw--Teitelboim  dilaton gravity with negative cosmological
constant admits a BF formulation based on the
$\mathfrak{sl}(2,\mathbb{R})$ algebra
\cite{Achucarro:1986uwr,Witten:1988hc}. The action is
\begin{equation}
S[\mathcal{X},\mathcal{A}]
=
\frac{k}{4\pi}
\int_{\mathcal{M}}
\mathfrak{tr}
\left(
\mathcal{X}\mathcal{F}
\right)
+S_{\rm bdy},
\end{equation}
where
\[
\mathcal{F}
=
d\mathcal{A}
+
\mathcal{A}\wedge\mathcal{A},
\]
$\mathcal{A}$ is an $\mathfrak{sl}(2,\mathbb{R})$ connection and
$\mathcal{X}$ is an adjoint-valued dilaton field. The generators
$\Lt_i$ ($i=0,\pm1$) satisfy
\begin{equation}
[\Lt_i,\Lt_j]
=
(i-j)\Lt_{i+j},
\end{equation}
with invariant bilinear form
\[
\mathfrak{tr}(\Lt_{\mp1}\Lt_{\pm1})=-2,
\qquad
\mathfrak{tr}(\Lt_0\Lt_0)=-\frac12 .
\]

The BF theory is invariant under
\begin{equation}
\delta_\lambda \mathcal{A}
=
d\lambda+[\mathcal{A},\lambda],
\qquad
\delta_\lambda \mathcal{X}
=
[\mathcal{X},\lambda],
\label{sec:inv}
\end{equation}
leading to the equations of motion
\begin{equation}
\mathcal{F}=0,
\qquad
d\mathcal{X}
+
[\mathcal{A},\mathcal{X}]
=
0.
\label{eq:BFeom}
\end{equation}
These equations imply that the connection is locally flat and that the adjoint field is covariantly constant along the gauge orbit, reflecting the topological nature of the theory~\cite{Witten:1991zz,Ikeda:1993fh,Schaller:1994es}.
Using radial gauge,
\begin{equation}
\mathcal{A}
=
b^{-1}ab+b^{-1}db,
\qquad
\mathcal{X}
=
b^{-1}xb,
\qquad
b(\rho)
=
e^{\rho \Lt_0},
\end{equation}
the reduced fields $a=a_t(t)\,dt$ and $x=x(t)$ become independent of
the radial coordinate. Imposing affine AdS$_2$ boundary conditions,
the connection determines the asymptotic geometry while the dilaton
field organizes the admissible residual gauge transformations. The
resulting asymptotic symmetry structure provides the holographic
dictionary underlying JT gravity and serves as the starting point for
its higher-spin extensions. 

Our analysis focuses on conformal boundary conditions for
$\mathfrak{sl}(2,\mathbb{R})$ JT dilaton gravity. Following the
Drinfeld--Sokolov reduction of the BF formulation
\cite{Grumiller:2016pqb,Ozer:2025bpb,Ozer:2017dwk,Ozer:2019nkv,Ozer:2021wtx,Ozer:2024ovo},
the asymptotically AdS$_2$ configuration is characterized by a single
boundary function $\mathcal{L}(t)$ appearing in the reduced
connection. The corresponding metric takes the Fefferman--Graham form
\begin{equation}
ds^2
=
d\rho^2
+
\left(
e^{2\rho}
+
2\mathcal{L}(t)
+
\mathcal{O}(e^{-2\rho})
\right)dt^2 .
\end{equation}
The corresponding dilaton configuration is given by
\begin{equation}
\label{AdSDilaton}
\mathcal{X}
	=
	e^{\rho}\mathcal{X}^{+}
	+
	e^{-\rho}\mathcal{X}^{-}.
\end{equation}

These boundary conditions provide the starting point for the affine
$\mathfrak{sl}(2,\mathbb{R})$ asymptotic symmetry analysis developed
below.

The equations of motion (\ref{eq:BFeom}) imply that BF gravity possesses no local propagating bulk degrees of freedom. Since the gauge connection is locally flat, it can always be written as a pure gauge configuration in a sufficiently small neighborhood. Similarly, the covariant constancy condition constrains the adjoint field to evolve along the corresponding gauge orbit. Consequently, the local bulk dynamics is entirely determined by gauge redundancy, and physical observables are associated instead with global structures, boundary conditions, and nontrivial holonomies. This characteristic feature distinguishes BF gravity from conventional local field theories and underlies the prominent role played by boundary phase spaces and asymptotic symmetry structures in low-dimensional gravitational systems~\cite{Witten:1991zz,Schaller:1994es,Gonzalez:2018enk}.


A particularly important realization of two-dimensional BF gravity is provided by Jackiw--Teitelboim gravity. In this case, the gauge algebra is chosen to be $\mathfrak{sl}(2,\mathbb{R})$, and the adjoint field is identified with the dilaton multiplet. The BF equations of motion then reproduce the geometric constraints of JT gravity, establishing an exact correspondence between the gauge-theoretic and geometric formulations of the theory. Within this framework, the metric and dilaton arise as derived quantities constructed from the underlying gauge variables, while the fundamental dynamics remains encoded in the flat connection and the adjoint field. This reformulation has proven particularly useful for the analysis of asymptotic symmetries, boundary conditions, and the emergence of effective boundary dynamics in AdS$_2$ gravity~\cite{Jackiw:1984je,Teitelboim:1983ux,Witten:1991zz,Ikeda:1993fh,Gonzalez:2018enk}.


Although the bulk dynamics of BF gravity is entirely constrained by the flatness and covariant constancy conditions, nontrivial physical structures arise once boundaries are introduced. In the presence of a boundary, not all gauge transformations are pure redundancies; a subset survives as residual transformations preserving the chosen boundary conditions. These residual symmetries act nontrivially on the boundary data and generate the associated asymptotic phase space. Consequently, the physical content of the theory is encoded not in local bulk excitations but in the structure of the boundary phase space together with its symmetry algebra. The precise form of this reduced phase space depends on the choice of boundary conditions and symmetry reductions, providing the starting point for the emergence of effective boundary theories such as the Schwarzian theory and its generalizations~\cite{Almheiri:2014cka,Maldacena:2016upp,Gonzalez:2018enk}.
\section{From BF Theory to the Schwarzian}
\label{sec:schwarzian}

The asymptotic phase space obtained from BF gravity is generally larger than the effective boundary theories that emerge from it. In particular, the residual symmetry algebra associated with a given set of boundary conditions typically contains more degrees of freedom than are required to describe the low-energy boundary dynamics. To extract physically relevant sectors, one must therefore impose additional constraints or reductions on the boundary phase space. Such reductions preserve the essential symmetry structure while eliminating redundant boundary degrees of freedom, thereby leading to simpler effective descriptions. From this perspective, boundary theories should be viewed not as fundamental objects but as reduced realizations of a more general asymptotic phase space endowed with residual gauge symmetries~\cite{Drinfeld:1984qv,Balog:1990mu,deBoer:1993iz,Gonzalez:2018enk}.

Among the various reductions that may be imposed on the boundary phase space, the Drinfeld--Sokolov reduction occupies a distinguished role. Originally developed in the study of integrable systems and Hamiltonian reductions of affine current algebras, it provides a systematic procedure for selecting a reduced set of boundary degrees of freedom while preserving a nontrivial symmetry structure~\cite{Drinfeld:1984qv,Balog:1990mu,deBoer:1993iz,Feher:1991bn}. In gravitational applications, the Drinfeld--Sokolov reduction can be understood as a particular choice of boundary constraints that transforms the original gauge-theoretic description into a reduced boundary theory. As a consequence, the infinite-dimensional structure of the asymptotic phase space becomes encoded in a smaller set of gauge-invariant quantities, which subsequently acquire the interpretation of physical boundary degrees of freedom.

For the case of $\mathfrak{sl}(2,\mathbb{R})$, the Drinfeld--Sokolov reduction is conveniently implemented through a highest-weight gauge for the boundary connection. In this gauge, the residual boundary degrees of freedom are encoded in a single state-dependent function, while the remaining components are fixed by the reduction conditions. The boundary connection can then be brought to a canonical form in which the asymptotic phase space is parameterized by a reduced set of variables rather than by the full affine current algebra. This gauge choice plays a central role in establishing the connection between BF gravity and Schwarzian dynamics, since it isolates the boundary data that survives the reduction and subsequently acquires a geometric interpretation~\cite{Drinfeld:1984qv,Balog:1990mu,deBoer:1993iz,Gonzalez:2018enk}.

We impose Drinfeld--Sokolov conformal boundary conditions on the
$\mathfrak{sl}(2,\mathbb{R})$ connection following the standard
Brown--Henneaux-type reduction
\cite{Brown:1986nw,Perez:2016vqo,Cardenas:2021vwo} with dilaton $x$ take the highest-weight
form
\begin{align}
a_\tau
&=
\Lt_{+1}
-
\mathcal{L} ~\Lt_{-1},
\label{eq:DSgauge}
\\
x
&=
\mathcal{X}\Lt_{1}
-\mathcal{X}'\Lt_{0}
-
\left(
\mathcal{L}\mathcal{X}
-
\frac{\mathcal{X}''}{2}
\right)\Lt_{-1}.
\end{align}

The residual gauge transformations are generated by a single function
$\epsilon$,
\begin{equation}
\lambda
=
b^{-1}
\left[
\epsilon\Lt_{1}
-
\epsilon'\Lt_{0}
-
\left(
\mathcal{L}\epsilon
-
\frac{\epsilon''}{2}
\right)\Lt_{-1}
\right]
b ,
\end{equation}
The highest-weight gauge does not completely eliminate gauge freedom. Instead, a restricted class of gauge transformations survives and preserves the form of the reduced connection. These residual transformations act nontrivially on the remaining state-dependent function and therefore generate the symmetry structure of the reduced phase space. Remarkably, the resulting transformation law coincides with the classical Virasoro coadjoint action, revealing that the reduced boundary dynamics is governed by a conformal symmetry algebra rather than by the original affine gauge algebra. In this way, the Drinfeld--Sokolov reduction converts the gauge-theoretic boundary description into a Virasoro-type phase space characterized by two dynamical functions ~\cite{Drinfeld:1984qv,deBoer:1993iz,Gonzalez:2018enk}.
\begin{align}
\delta_{\lambda}\mathcal{L}
&=-
\frac{1}{2}\epsilon'''
+
\epsilon\mathcal{L}'
+
2\mathcal{L}\epsilon',
\\
\delta_{\lambda}\mathcal{X}
&=
\epsilon\mathcal{X}'
-
\epsilon'\mathcal{X}.
\end{align}
The field $\mathcal{L}$ therefore transforms as a Virasoro
coadjoint vector, while $\mathcal{X}$ transforms as a conformal
field of weight $-1$.

The Virasoro structure obtained after the Drinfeld--Sokolov reduction admits a natural geometric interpretation in terms of boundary reparametrizations. Rather than describing the reduced phase space through the state-dependent function appearing in the highest-weight connection, one may equivalently parameterize the same degrees of freedom by a boundary diffeomorphism. In this description, the Virasoro variable is expressed through the Schwarzian derivative of the reparametrization field, establishing a direct correspondence between the reduced BF phase space and Schwarzian geometry. Consequently, the Schwarzian derivative appears not as an additional dynamical input but as a geometric quantity naturally associated with the reduced Virasoro orbit arising from the Drinfeld--Sokolov reduction~\cite{Maldacena:2016upp,Stanford:2017thb,Mertens:2017mtv,Alekseev:1988ce}.


The relation between the Drinfeld--Sokolov variable
$\mathcal L(\tau)$ and the Schwarzian derivative can be made explicit
by rewriting the reduced $\mathfrak{sl}(2,\mathbb R)$ problem as a
second-order differential equation. In the fundamental representation,
the highest-weight connection (\ref{eq:DSgauge}) defines the linear
system
\begin{equation}
\left(\partial_\tau+a_\tau\right)\Psi=0 ,
\label{eq:linear_system_sl2}
\end{equation}
for a two-component vector $\Psi$. Eliminating one component gives a
second-order equation of Hill type,
\begin{equation}
\psi''-\mathcal L(\tau)\psi=0 .
\label{eq:HillEquation}
\end{equation}
Thus the single function $\mathcal L(\tau)$ appearing in the
Drinfeld--Sokolov connection plays the role of the projective
connection associated with the reduced boundary phase space.

Let $\psi_1$ and $\psi_2$ be two independent solutions of
(\ref{eq:HillEquation}), and define the boundary reparametrization
field by their ratio
\begin{equation}
f(\tau)=\frac{\psi_1(\tau)}{\psi_2(\tau)} .
\label{eq:ratio_solutions}
\end{equation}
The Wronskian
\begin{equation}
W=\psi_1'\psi_2-\psi_1\psi_2'
\label{eq:Wronskian}
\end{equation}
is constant as a consequence of (\ref{eq:HillEquation}). From
(\ref{eq:ratio_solutions}) one obtains
\begin{equation}
f'
=
\frac{W}{\psi_2^2},
\qquad
\frac{f''}{f'}
=
-2\,\frac{\psi_2'}{\psi_2}.
\label{eq:fprime_ratio}
\end{equation}
Differentiating once more gives
\begin{equation}
\left(\frac{f''}{f'}\right)'
=
-2
\left(
\frac{\psi_2''}{\psi_2}
-
\frac{\psi_2'^2}{\psi_2^2}
\right).
\label{eq:ratio_derivative}
\end{equation}
Using the Hill equation, $\psi_2''=\mathcal L\psi_2$, this becomes
\begin{equation}
\left(\frac{f''}{f'}\right)'
=
-2\mathcal L
+
2\left(
\frac{\psi_2'}{\psi_2}
\right)^2 .
\label{eq:ratio_derivative_L}
\end{equation}
The Schwarzian derivative can alternatively be written as

\begin{equation}
\{f,\tau\}
=
\left(
\frac{f''}{f'}
\right)'
-
\frac12
\left(
\frac{f''}{f'}
\right)^2 .
\label{eq:Schwarzian_alt}
\end{equation}
Substituting (\ref{eq:fprime_ratio}) and
(\ref{eq:ratio_derivative_L}) into (\ref{eq:Schwarzian_alt}) yields
\begin{equation}
\{f,\tau\}
=
-2\mathcal L(\tau).
\label{eq:Schwarzian_projective_relation}
\end{equation}
Consequently, the reduced BF phase-space variable
$\mathcal L(\tau)$ is directly related to the Schwarzian derivative as
\begin{equation}
\mathcal L(\tau)
=
-\frac12\,\{f,\tau\}.
\label{eq:Lsch}
\end{equation}
where
\[
\{f,\tau\}
=
\frac{f'''(\tau)}{f'(\tau)}
-\frac32
\left(
\frac{f''(\tau)}{f'(\tau)}
\right)^2
\]
denotes the Schwarzian derivative.
This is the precise sense in which the Schwarzian derivative emerges
from the projective structure defined by the reduced
$\mathfrak{sl}(2,\mathbb R)$ BF connection.

Substituting the Schwarzian parametrization into the reduced boundary description yields the familiar Schwarzian action governing the effective boundary dynamics of the theory. In this formulation, the dynamics is entirely encoded in the boundary reparametrization field, while the original gauge-theoretic degrees of freedom are absorbed into the reduced geometric description. The resulting theory captures the low-energy sector of the reduced BF phase space and provides an effective representation of the underlying Virasoro orbit structure. The corresponding boundary action takes the form
\begin{equation}
S_{\rm Sch}[f]
=
-C
\int d\tau
\{f,\tau\}
\label{eq:SchwarzianAction}
\end{equation}

where $C$ is an effective coupling constant. The action (\ref{eq:SchwarzianAction}) provides the standard low-energy description of JT boundary dynamics~\cite{Maldacena:2016upp,Stanford:2017thb,Mertens:2017mtv,Engelsoy:2016xyb}. Within the present framework, however, it is interpreted as the endpoint of a sequence of reductions beginning with BF gravity and proceeding through the Drinfeld--Sokolov construction. The natural question is whether an analogous construction exists for higher-rank gauge algebras. We now turn to the $\mathfrak{sl}(3,\mathbb R)$ BF theory and show that its reduced phase space gives rise to a generalized Schwarzian structure governed by the second and third Wilczynski invariants.

\section{Generalization to $\mathfrak{sl}(3,\mathbb{R})$}
\label{sec:sl3}
The simplest higher-spin extension of JT dilaton gravity is obtained by
enlarging the underlying gauge algebra from
$\mathfrak{sl}(2,\mathbb{R})$ to $\mathfrak{sl}(3,\mathbb{R})$. This
extension introduces an additional spin-3 sector while preserving the
BF-theoretic structure of the theory. Consequently, the gravitational
multiplet associated with the $\mathfrak{sl}(2,\mathbb{R})$ generators
is supplemented by new degrees of freedom encoded in higher-spin
fields.

Under the principal embedding of $\mathfrak{sl}(2,\mathbb{R})$ into
$\mathfrak{sl}(3,\mathbb{R})$, the algebra decomposes into the
$\mathfrak{sl}(2,\mathbb{R})$ generators $\Lt_i$
($i=-1,0,+1$) and a quintet of spin-3 generators $\Wt_m$
($m=-2,-1,0,+1,+2$), whose non-vanishing commutation relations read
\begin{align}
[\Lt_i,\Lt_j]
&=
(i-j)\Lt_{i+j},
\\
[\Lt_i,\Wt_m]
&=
(2i-m)\Wt_{i+m},
\\
[\Wt_n,\Wt_m]
&=
\sigma (n-m)
\left(
2n^2-nm+2m^2-8
\right)
\Lt_{n+m},
\end{align}
where the constant $\sigma$ fixes the normalization of the spin-3
sector.

The corresponding invariant bilinear form is chosen according to
\begin{equation}
\tr(\Lt_{\mp1}\Lt_{\pm1})
= -2,\tr(\Lt_0\Lt_0)
-1,
\end{equation}
and
\begin{equation}
\tr(\Wt_{\mp2}\Wt_{\pm2})
= -4\,\tr(\Wt_{\mp1}\Wt_{\pm1})
= 6\,\tr(\Wt_0\Wt_0)=
-48\sigma .
\end{equation}

The algebraic structure introduced above provides the foundation for
the $\mathfrak{sl}(3,\mathbb{R})$ BF formulation, from which the
higher-spin connection, generalized dilaton multiplet, and their
associated asymptotic symmetry structure will be derived in the
following subsections.

The construction described in Section~\ref{sec:schwarzian} admits a
natural higher-rank generalization. In the $\mathfrak{sl}(2,\mathbb{R})$
case, the Drinfeld--Sokolov reduction leads to a second-order
projective equation, whose projective invariant is the Schwarzian
derivative. For $\mathfrak{sl}(3,\mathbb{R})$, the reduced connection
contains two independent state-dependent functions. Therefore, the
reduced boundary dynamics is no longer characterized by a single
Schwarzian derivative, but by two independent projective invariants.

We consider the $\mathfrak{sl}(3,\mathbb{R})$ BF theory with connection
$\mathcal{A}$ and adjoint dilaton field $\mathcal{X}$,

\begin{equation}
\mathcal{A}=A^{A}\Tt_{A},
\qquad
\mathcal{X}=X^{A}\Tt_{A},
\qquad
\Tt_{A}\in\mathfrak{sl}(3,\mathbb{R}) .
\label{eq:sl3_fields}
\end{equation}
The BF action keeps the same form as in Eq.~\eqref{eq:BFeom}.
The difference from the $\mathfrak{sl}(2,\mathbb{R})$ case lies not in
the BF structure itself, but in the enlarged gauge algebra and in the
corresponding Drinfeld--Sokolov reduction.

The reduced boundary connections may then be chosen
in the highest-weight Drinfeld--Sokolov form
\begin{align}
a_\tau
&=
\Lt_{+1}
- \mathcal{L}\,\Lt_{-1}
- \mathcal{W}\,\Wt_{-2},
\label{at_sl3}
\\[2mm]
x
&=
\mathcal{X}\,\Lt_{1}
-
\mathcal{X}'\,\Lt_{0}
+
\mathcal{Y}\,\Wt_{2}
-
\mathcal{Y}'\,\Wt_{1}
\nonumber\\[1mm]
&\quad
+
\left(
- \mathcal{L}\mathcal{X}
+
\frac{\mathcal{X}''}{2}
+
24\,\,\mathcal{W}\mathcal{Y}
\right)\Lt_{-1}
\nonumber\\[1mm]
&\quad
+
\left(
- \mathcal{W}\mathcal{X}
-
\frac{7}{12}\mathcal{Y}'\mathcal{L}'
-
\frac{2}{3} \mathcal{L}\mathcal{Y}''
+
\frac{\mathcal{Y}^{(4)}}{24}
+
\mathcal{Y}\mathcal{L}^{2}
-
\frac{1}{6} \mathcal{Y}\mathcal{L}''
\right)\Wt_{-2}
\nonumber\\[1mm]
&\quad
+
\left(
\frac{5}{3} \mathcal{L}\mathcal{Y}'
-
\frac{\mathcal{Y}^{(3)}}{6}
+
\frac{2}{3} \mathcal{Y}\mathcal{L}'
\right)\Wt_{-1}
\nonumber\\[1mm]
&\quad
+
\left(
\frac{\mathcal{Y}''}{2}
-
2 \mathcal{Y}\mathcal{L}
\right)\Wt_{0}.
\label{x_sl3}
\end{align}
Here $\mathcal L=\mathcal W_2(\tau)$ is the spin-2 component and
$\mathcal W=\mathcal W_3(\tau)$ is the spin-3 component of the reduced boundary
data. The first one generalizes the projective connection
$\mathcal L(\tau)$ of the $\mathfrak{sl}(2,\mathbb{R})$ case, whereas
the second one is a genuinely higher-spin contribution.

The residual gauge transformations are generated by two single functions $\epsilon$ and $\eta$,
\begin{align}
\lambda
&=
b^{-1}
\bigg[
\epsilon\,\Lt_{1}
-
\epsilon'\,\Lt_{0}
+
\eta\,\Wt_{2}
-
\eta'\,\Wt_{1}
\nonumber\\[1mm]
&\quad
+
\left(
-\mathcal{L}\epsilon
+
\frac{\epsilon''}{2}
+
24\,\mathcal{W}\eta
\right)\Lt_{-1}
\nonumber\\[1mm]
&\quad
+
\left(
-\mathcal{W}\epsilon
-
\frac{7}{12}\eta'\mathcal{L}'
-
\frac{2}{3}\mathcal{L}\eta''
+
\frac{\eta^{(4)}}{24}
+
\eta\mathcal{L}^{2}
-
\frac{1}{6}\eta\mathcal{L}''
\right)\Wt_{-2}
\nonumber\\[1mm]
&\quad
+
\left(
\frac{5}{3}\mathcal{L}\eta'
-
\frac{\eta^{(3)}}{6}
+
\frac{2}{3}\eta\mathcal{L}'
\right)\Wt_{-1}
\nonumber\\[1mm]
&\quad
+
\left(
\frac{\eta''}{2}
-
2\eta\mathcal{L}
\right)\Wt_{0}
\bigg]
b .
\label{lambda_sl3}
\end{align}
By inserting this gauge parameter into the field transformation
equation given in (\ref{sec:inv}), we derive the corresponding infinitesimal gauge transformations:
\begin{align}
\delta_{\lambda}\mathcal{L}
&=
\frac{k}{4}\epsilon^{(3)}
+
2\mathcal{L}\epsilon'
+
\epsilon\mathcal{L}'
+
2\eta\mathcal{W}'
+
3\mathcal{W}\eta',
\label{eq:sl3_varL}
\\[2mm]
\delta_{\lambda}\mathcal{W}
&=
\epsilon\mathcal{W}'
+
3\mathcal{W}\epsilon'
+
\frac{32}{15k}
\left(
\mathcal{L}^{2}\eta'
+
\eta\mathcal{L}\mathcal{L}'
\right)
+
\frac{k}{120}\eta^{(5)}
\nonumber\\
&\quad
+
\frac{1}{15}
\left(
3\eta'\mathcal{L}''
+
5\eta''\mathcal{L}'
\right)
+
\frac{1}{10}
\left(
\eta\mathcal{L}^{(3)}
+
5\eta^{(3)}\mathcal{L}
\right),
\label{eq:sl3_varW}
\\[2mm]
\delta_{\lambda}\mathcal{X}
&=
\frac{1}{10}
\left(
\eta''\mathcal{Y}'
-
\eta'\mathcal{Y}''
\right)
-
\frac{1}{15}
\left(
\eta^{(3)}\mathcal{Y}
-
\eta\mathcal{Y}^{(3)}
\right)
\nonumber\\
&\quad
-
\frac{32}{15k}
\left(
\mathcal{Y}\mathcal{L}\eta'
-
\eta\mathcal{L}\mathcal{Y}'
\right)
-
\mathcal{X}\epsilon'
+
\epsilon\mathcal{X}',
\label{eq:sl3_varX}
\\[2mm]
\delta_{\lambda}\mathcal{Y}
&=
-\mathcal{X}\eta'
+
2\eta\mathcal{X}'
-
2\mathcal{Y}\epsilon'
+
\epsilon\mathcal{Y}' .
\label{eq:sl3_varY}
\end{align}
The spin-2 and spin-3 charges $\mathcal L=\mathcal W_2(\tau)$ and $\mathcal W=\mathcal W_3(\tau)$ have
conformal weights $\Delta=2$ and $\Delta=3$, while the corresponding dilaton fields $\mathcal{X}$ and $\mathcal{Y}$ possess weights
$-1$ and $-2$, respectively.

The analogue of the Hill equation is obtained by rewriting the
$\mathfrak{sl}(3,\mathbb{R})$ Drinfeld--Sokolov system in companion
form. Equivalently, one may encode the reduced connection in the
third-order differential equation
\begin{equation}
\psi'''
-
\mathcal W_2(\tau)\psi'
-
\mathcal W_3(\tau)\psi
=
0 .
\label{eq:sl3_third_order_ode}
\end{equation}

This equation is the direct higher-rank analogue of the second-order
equation
\begin{equation}
\psi''
-
\mathcal L(\tau)\psi
=
0
\label{eq:sl2_hill_again}
\end{equation}
encountered in the $\mathfrak{sl}(2,\mathbb{R})$ reduction. Thus the
transition from $\mathfrak{sl}(2,\mathbb{R})$ to
$\mathfrak{sl}(3,\mathbb{R})$ replaces the projective line data of a
second-order equation by the projective-plane data of a third-order
equation.

Let $\psi_1,\psi_2,\psi_3$ be three independent solutions of
Eq.~\eqref{eq:sl3_third_order_ode}. They define a projective curve in the two-dimensional projective space.
 In an affine chart this curve may be represented as
\begin{equation}
Y(\tau)
=
\begin{pmatrix}
1\\
f(\tau)\\
g(\tau)
\end{pmatrix}.
\label{eq:projective_curve_fg}
\end{equation}
The two functions $f(\tau)$ and $g(\tau)$ replace the single
reparametrization field appearing in the ordinary Schwarzian theory.
Consequently, the geometric data of the reduced
$\mathfrak{sl}(3,\mathbb{R})$ boundary theory are encoded in the
projective geometry of curves in the two-dimensional projective space.

The intrinsic invariants of such projective curves are the Wilczynski
invariants. In the present context, the second and third Wilczynski
invariants will be denoted by $I_2(\tau)$ and $I_3(\tau)$. They play
the role of the generalized Schwarzian data associated with the
$\mathfrak{sl}(3,\mathbb{R})$ BF reduction. In the next section, we
derive these invariants explicitly from a normalized projective lift and
show how they are related to the reduced fields
$\mathcal W_2(\tau)$ and $\mathcal W_3(\tau)$.

\section{Wilczynski Invariants from Flat BF Connections}
\label{sec:wilczynski}

The third-order differential equation obtained in
Section~\ref{sec:sl3} naturally defines a projective curve in the two-dimensional projective space. In an affine chart, this curve may be represented as

\begin{equation}
Y(\tau)
=
\begin{pmatrix}
1\\
f(\tau)\\
g(\tau)
\end{pmatrix}.
\label{eq:Yfg}
\end{equation}

The functions $f(\tau)$ and $g(\tau)$ provide local coordinates on the
projective curve. The relevant projective information is encoded not in
the choice of representative itself but in its normalized lift.


Define

\begin{equation}
\Delta
=
\det(Y,Y',Y'')
=
f'g''-f''g' .
\label{eq:Delta}
\end{equation}

Assuming $\Delta\neq0$, we introduce the normalized lift

\begin{equation}
X
=
\Delta^{-1/3}Y .
\label{eq:normalized_lift}
\end{equation}

A direct computation shows that

\begin{equation}
\det(X,X',X'')=1 .
\label{eq:unimodular_lift}
\end{equation}

This condition is the natural
$\mathfrak{sl}(3,\mathbb R)$ analogue of fixing the Wronskian in the
second-order $\mathfrak{sl}(2,\mathbb R)$ problem.

Since $X$, $X'$ and $X''$ form a basis of the three-dimensional solution space, the third
derivative must be expandable as

\begin{equation}
X'''
=
A(\tau)X
+
B(\tau)X'
+
C(\tau)X'' .
\label{eq:Xthird_general}
\end{equation}

Differentiating Eq.~\eqref{eq:unimodular_lift} yields

\begin{equation}
0
=
\frac{d}{d\tau}
\det(X,X',X'')
=
\det(X,X',X''').
\label{eq:det_derivative}
\end{equation}

Substituting Eq.~\eqref{eq:Xthird_general} into
Eq.~\eqref{eq:det_derivative} immediately implies

\begin{equation}
C(\tau)=0 .
\label{eq:Czero}
\end{equation}

Therefore,

\begin{equation}
X'''
=
I_3(\tau)X
+
I_2(\tau)X' ,
\label{eq:Wilczynski_closure}
\end{equation}

which defines the second and third Wilczynski invariants.


Introduce

\begin{equation}
\rho=\Delta^{-1/3},
\qquad
u=\frac{\Delta'}{\Delta},
\qquad
a=\frac{\rho'}{\rho}
=
-\frac13u .
\label{eq:rho_u}
\end{equation}

Since $X=\rho Y$, one finds

\begin{equation}
X'
=
\rho
\left(
Y'+aY
\right),
\label{eq:Xprime}
\end{equation}

\begin{equation}
X''
=
\rho
\left[
Y''
+
2aY'
+
(a'+a^2)Y
\right],
\label{eq:Xdoubleprime}
\end{equation}

and

\begin{equation}
X'''
=
\rho
\left[
Y'''
+
3aY''
+
3(a'+a^2)Y'
+
(a''+3aa'+a^3)Y
\right].
\label{eq:Xtripleprime}
\end{equation}


Since

\begin{equation}
Y'''
=
\begin{pmatrix}
0\\
f'''\\
g'''
\end{pmatrix},
\label{eq:Ythird}
\end{equation}

the vector $Y'''$ admits the decomposition

\begin{equation}
Y'''
=
qY'
+
rY'' .
\label{eq:Ythird_decomp}
\end{equation}

The coefficients satisfy

\begin{equation}
f'''
=
qf'
+
rf'',
\qquad
g'''
=
qg'
+
rg'' .
\label{eq:qr_system}
\end{equation}

Solving gives

\begin{equation}
q
=
\frac{f'''g''-f''g'''}
{\Delta},
\qquad
r
=
\frac{f'g'''-f'''g'}
{\Delta}.
\label{eq:q_r}
\end{equation}

Since

\begin{equation}
\Delta'
=
f'g'''
-
f'''g',
\label{eq:Deltaprime}
\end{equation}

one obtains

\begin{equation}
r
=
\frac{\Delta'}{\Delta}
=
u .
\label{eq:r_equals_u}
\end{equation}


Substituting Eq.~\eqref{eq:Ythird_decomp} into
Eq.~\eqref{eq:Xtripleprime} yields

\begin{equation}
X'''
=
\rho
\left[
(a''+3aa'+a^3)Y
+
(q+3a'+3a^2)Y'
+
(r+3a)Y''
\right].
\label{eq:Xthird_expanded}
\end{equation}

Using Eqs.~\eqref{eq:rho_u} and \eqref{eq:r_equals_u}, we find

\begin{equation}
r+3a=0 .
\label{eq:Ydouble_cancel}
\end{equation}

Thus the $Y''$ component disappears identically, as required by the
unimodular normalization.


Comparing Eq.~\eqref{eq:Xthird_expanded} with
Eq.~\eqref{eq:Wilczynski_closure} and using
Eq.~\eqref{eq:Xprime}, one obtains

\begin{equation}
I_2
=
q+3a'+3a^2 .
\label{eq:I2_first}
\end{equation}

Using Eq.~\eqref{eq:rho_u}, this becomes

\begin{equation}
I_2
=
q-u'
+\frac13u^2 .
\label{eq:I2_qu}
\end{equation}

Hence

\begin{equation}
I_2
=
\frac{f'''g''-f''g'''}
{\Delta}
-
\left(
\frac{\Delta'}{\Delta}
\right)'
+
\frac13
\left(
\frac{\Delta'}{\Delta}
\right)^2 .
\label{eq:I2_explicit}
\end{equation}


Comparing the coefficients of $Y$ in
Eq.~\eqref{eq:Xthird_expanded} and
Eq.~\eqref{eq:Wilczynski_closure} yields

\begin{equation}
I_3+aI_2
=
a''+3aa'+a^3 .
\label{eq:I3_relation}
\end{equation}

Therefore

\begin{equation}
I_3
=
a''+3aa'+a^3-aI_2 .
\label{eq:I3_a}
\end{equation}

Substituting Eq.~\eqref{eq:rho_u} and Eq.~\eqref{eq:I2_qu}, one obtains

\begin{equation}
I_3
=
\frac13uq
-\frac13u''
+\frac{2}{27}u^3 .
\label{eq:I3_qu}
\end{equation}

or equivalently

\begin{equation}
I_3
=
\frac13
\frac{\Delta'}{\Delta}
\frac{f'''g''-f''g'''}
{\Delta}
-\frac13
\left(
\frac{\Delta'}{\Delta}
\right)''
+\frac{2}{27}
\left(
\frac{\Delta'}{\Delta}
\right)^3 .
\label{eq:I3_explicit}
\end{equation}
\section{Generalized Schwarzian Actions}
\label{sec:gsch}

The ordinary Schwarzian action can be interpreted as the integral of
the projective invariant associated with the reduced
$\mathfrak{sl}(2,\mathbb R)$ Drinfeld--Sokolov connection. In the
$\mathfrak{sl}(3,\mathbb R)$ case, the corresponding reduced
projective data are given by the pair of Wilczynski invariants
$I_2(\tau)$ and $I_3(\tau)$ derived in
Section~\ref{sec:wilczynski}. This motivates the generalized
Schwarzian functional
\begin{equation}
S_{\rm gSch}[f,g]
=
\int d\tau\,
\left[
\alpha I_2(\tau)
+
\beta I_3(\tau)
\right],
\label{eq:gSch_action_I2I3}
\end{equation}
where $\alpha$ and $\beta$ are coupling constants associated with the
spin-2 and spin-3 sectors, respectively.


Using Eqs.~\eqref{eq:I2_qu} and \eqref{eq:I3_qu}, the action can be
written in terms of the auxiliary variables
\begin{equation}
u=\frac{\Delta'}{\Delta},
\qquad
q=
\frac{f'''g''-f''g'''}{\Delta},
\qquad
\Delta=f'g''-f''g',
\label{eq:uqDelta}
\end{equation}
as
\begin{equation}
S_{\rm gSch}[f,g]
=
\int d\tau
\left[
\alpha
\left(
q-u'
+\frac13u^2
\right)
+
\beta
\left(
\frac13uq
-\frac13u''
+\frac{2}{27}u^3
\right)
\right].
\label{eq:gSch_action_uq}
\end{equation}
Up to total derivatives, this may be simplified to
\begin{equation}
S_{\rm gSch}[f,g]
\simeq
\int d\tau
\left[
\alpha
\left(
q+\frac13u^2
\right)
+
\beta
\left(
\frac13uq
+
\frac{2}{27}u^3
\right)
\right],
\label{eq:gSch_action_reduced}
\end{equation}
where $\simeq$ denotes equality modulo boundary terms.


The variation of the action is obtained from the variations of
$\Delta$, $u$, and $q$. First,
\begin{equation}
\delta\Delta
=
\delta f'\,g''
+
f'\,\delta g''
-
\delta f''\,g'
-
f''\,\delta g' .
\label{eq:deltaDelta}
\end{equation}
Therefore,
\begin{equation}
\delta u
=
\delta\left(\frac{\Delta'}{\Delta}\right)
=
\frac{(\delta\Delta)'}{\Delta}
-
\frac{\Delta'}{\Delta^2}\delta\Delta
=
\left(
\frac{\delta\Delta}{\Delta}
\right)' .
\label{eq:deltau}
\end{equation}
Writing
\begin{equation}
q=\frac{N}{\Delta},
\qquad
N=f'''g''-f''g''',
\label{eq:Ndef}
\end{equation}
one also finds
\begin{equation}
\delta q
=
\frac{\delta N}{\Delta}
-
\frac{N}{\Delta^2}\delta\Delta ,
\label{eq:deltaq}
\end{equation}
with
\begin{equation}
\delta N
=
\delta f'''\,g''
+
f'''\,\delta g''
-
\delta f''\,g'''
-
f''\,\delta g''' .
\label{eq:deltaN}
\end{equation}


Using Eq.~\eqref{eq:gSch_action_uq}, the first variation takes the
form
\begin{equation}
\delta S_{\rm gSch}
=
\int d\tau
\left[
\mathcal P\,\delta q
+
\mathcal Q\,\delta u
+
\mathcal R\,\delta u'
+
\mathcal T\,\delta u''
\right],
\label{eq:deltaS_general}
\end{equation}
where
\begin{equation}
\mathcal P
=
\alpha+\frac{\beta}{3}u,
\qquad
\mathcal Q
=
\frac{2\alpha}{3}u+\frac{\beta}{3}q+\frac{2\beta}{9}u^2,
\qquad
\mathcal R=-\alpha,
\qquad
\mathcal T=-\frac{\beta}{3}.
\label{eq:PQRT}
\end{equation}
After integrating by parts, the terms involving $\delta u'$ and
$\delta u''$ may be absorbed into the coefficient of $\delta u$.
Equivalently,
\begin{equation}
\delta S_{\rm gSch}
=
\int d\tau
\left[
\mathcal P\,\delta q
+
\mathcal U\,\delta u
\right]
+
\text{boundary terms},
\label{eq:deltaS_PU}
\end{equation}
where
\begin{equation}
\mathcal U
=
\mathcal Q
-
\mathcal R'
+
\mathcal T''
=
\frac{2\alpha}{3}u
+
\frac{\beta}{3}q
+
\frac{2\beta}{9}u^2 .
\label{eq:Udef}
\end{equation}
Here $\mathcal R'=0$ and $\mathcal T''=0$ because $\alpha$ and
$\beta$ are taken to be constants.


Substituting Eqs.~\eqref{eq:deltau} and \eqref{eq:deltaq} into
Eq.~\eqref{eq:deltaS_PU}, one obtains
\begin{equation}
\delta S_{\rm gSch}
=
\int d\tau
\left[
\mathcal P
\left(
\frac{\delta N}{\Delta}
-
\frac{N}{\Delta^2}\delta\Delta
\right)
+
\mathcal U
\left(
\frac{\delta\Delta}{\Delta}
\right)'
\right]
+
\text{boundary terms}.
\label{eq:deltaS_DeltaN}
\end{equation}
Integrating the last term by parts gives
\begin{equation}
\delta S_{\rm gSch}
=
\int d\tau
\left[
\frac{\mathcal P}{\Delta}\delta N
-
\left(
\frac{\mathcal P N}{\Delta^2}
+
\left(
\frac{\mathcal U}{\Delta}
\right)'
\right)
\delta\Delta
\right]
+
\text{boundary terms}.
\label{eq:deltaS_compact}
\end{equation}
It is useful to define
\begin{equation}
A_0
=
\frac{\mathcal P}{\Delta},
\qquad
B_0
=
-
\frac{\mathcal P N}{\Delta^2}
-
\left(
\frac{\mathcal U}{\Delta}
\right)' .
\label{eq:A0B0}
\end{equation}
Then
\begin{equation}
\delta S_{\rm gSch}
=
\int d\tau
\left[
A_0\,\delta N
+
B_0\,\delta\Delta
\right]
+
\text{boundary terms}.
\label{eq:deltaS_A0B0}
\end{equation}


Using Eqs.~\eqref{eq:deltaDelta} and \eqref{eq:deltaN}, the variation
can be written in the Euler--Lagrange form
\begin{equation}
\delta S_{\rm gSch}
=
\int d\tau
\left[
E_f\,\delta f
+
E_g\,\delta g
\right]
+
\text{boundary terms}.
\label{eq:deltaS_EL}
\end{equation}
The two equations of motion are therefore
\begin{equation}
E_f=0,
\qquad
E_g=0.
\label{eq:gSch_EOM}
\end{equation}
Explicitly, the $f$-equation is
\begin{equation}
E_f
=
-\frac{d^3}{d\tau^3}
\left(
A_0 g''
\right)
+
\frac{d^2}{d\tau^2}
\left(
A_0 g'''
+
B_0 g'
\right)
-
\frac{d}{d\tau}
\left(
B_0 g''
\right)
=0,
\label{eq:Ef}
\end{equation}
while the $g$-equation is
\begin{equation}
E_g
=
-\frac{d^2}{d\tau^2}
\left(
A_0 f'''
+
B_0 f'
\right)
+
\frac{d^3}{d\tau^3}
\left(
A_0 f''
\right)
+
\frac{d}{d\tau}
\left(
B_0 f''
\right)
=0 .
\label{eq:Eg}
\end{equation}
Equations~\eqref{eq:Ef} and \eqref{eq:Eg} are the Euler--Lagrange
equations of the $\mathfrak{sl}(3,\mathbb R)$ generalized Schwarzian
functional.


The structure of Eqs.~\eqref{eq:Ef} and \eqref{eq:Eg} reflects the
higher-rank nature of the generalized Schwarzian theory. Unlike the
ordinary Schwarzian action, whose dynamics is governed by a single
boundary reparametrization field, the $\mathfrak{sl}(3,\mathbb R)$
theory involves two projective coordinates and therefore two coupled
Euler--Lagrange equations. The appearance of higher derivatives follows
from the fact that the Wilczynski invariants depend on derivatives of
$f$ and $g$ up to third order. This provides a concrete variational
realization of the generalized Schwarzian dynamics associated with the
reduced $\mathfrak{sl}(3,\mathbb R)$ BF phase space.
\section{Dilaton Sectors and Stabilizer Structures}
\label{sec:dilaton_stabilizers}
The previous sections focused on the connection sector and its
associated projective differential equations, leading to the
Schwarzian and generalized Schwarzian actions. BF gravity, however,
contains not only the gauge connection but also an adjoint-valued
dilaton multiplet. It is therefore natural to ask whether the
dilaton sector exhibits a parallel geometric structure. As we show
below, the spin-2 dilaton is governed by a third-order stabilizer
equation related to the Hill system, while the spin-3 dilaton sector
suggests a natural higher-spin extension of this construction.

For the $\mathfrak{sl}(2,\mathbb{R})$ theory, the dilaton field is
chosen in the same form as the residual gauge parameter preserving
the Drinfeld--Sokolov connection,
\begin{equation}
x
=
yL_{1}
-
y'L_{0}
+
\left(
\mathcal{L}y
+
\frac12 y''
\right)L_{-1}.
\label{eq:sl2_dilaton}
\end{equation}
Since the dilaton is covariantly constant on shell, it acts as a
stabilizer of the background connection. Requiring
\begin{equation}
\delta_y\mathcal{L}=0,
\end{equation}
leads to the third-order equation
\begin{equation}
\left(
\partial_\tau^3
+
4\mathcal{L}\partial_\tau
+
2\mathcal{L}'
\right)y
=
0.
\label{eq:sl2_stabilizer}
\end{equation}

Let $\psi_1$ and $\psi_2$ denote two independent solutions of the
Hill equation derived in the previous section. Then the general local
solution of Eq.~\eqref{eq:sl2_stabilizer} can be written as
\begin{equation}
y
=
c_1\psi_1^2
+
c_2\psi_1\psi_2
+
c_3\psi_2^2 .
\label{eq:sl2_dilaton_solution}
\end{equation}
Thus, the spin-2 dilaton sector does not introduce an independent
projective structure. Rather, it is completely determined by
quadratic combinations of the same Hill solutions that generate the
Schwarzian degree of freedom.

The structure of Eq.~\eqref{eq:sl2_dilaton_solution} is closely related
to the symmetric-square construction encountered in projective
differential geometry, where certain third-order differential
equations arise as symmetric-square lifts of underlying second-order
Hill-type systems \cite{OvsienkoTabachnikov}. This observation provides a
geometric interpretation of the dilaton stabilizer equation in terms
of the same projective structure that underlies the Schwarzian
sector. In particular, the dilaton stabilizer equation may be viewed
as a projective-geometric companion of the Schwarzian dynamics rather
than as an independent dynamical structure.

The same logic motivates an analogous interpretation in the
$\mathfrak{sl}(3,\mathbb{R})$ theory. There, the reduced connection
is governed by the third-order Wilczynski equation discussed in the
previous sections, while the corresponding dilaton multiplet is
described by two functions $(\mathcal X,\mathcal Y)$. The associated
stabilizer conditions take the form
\begin{equation}
\delta_{\mathcal X,\mathcal Y}\mathcal L=0,
\qquad
\delta_{\mathcal X,\mathcal Y}\mathcal W=0,
\label{eq:sl3_stabilizer}
\end{equation}
which constitute the natural $W_3$ analogue of the Virasoro
stabilizer equation.

Let $\psi_1$, $\psi_2$ and $\psi_3$ denote three independent
solutions of the Wilczynski equation. Motivated by the spin-2
construction, it is natural to conjecture that the spin-3 dilaton
sector may similarly be organized by quadratic combinations of these
solutions,
\begin{equation}
\psi_i\psi_j,
\qquad
i,j=1,2,3.
\label{eq:sl3_quadratic}
\end{equation}
Interestingly, the number of independent quadratic combinations of
three Wilczynski solutions coincides with the number of independent
coefficients appearing in the spin-2/spin-3 dilaton multiplet.

The spin-2 analysis suggests that the corresponding
dilaton sector is naturally related to composite
structures built from the solutions of the underlying
projective differential equation. Motivated by this
observation, it is tempting to expect a similar pattern
in the spin-3 theory, where the dilaton multiplet may
be related to quadratic combinations of Wilczynski
solutions. A complete derivation of this relation,
however, remains an interesting open problem.

\section{Holonomies, Monodromies and Boundary Data}
\label{sec:monodromy}


The generalized Schwarzian theory derived in the previous sections
admits a distinguished class of constant saddle configurations. These
are characterized by constant values of the Wilczynski invariants,

\begin{equation}
I_2(\tau)=I_{2,0},
\qquad
I_3(\tau)=I_{3,0}.
\label{eq:constant_I2I3}
\end{equation}

Such configurations represent stationary points of the reduced phase
space and provide the natural starting point for the analysis of the
global boundary structure.


For constant Wilczynski data, the companion connection becomes

\begin{equation}
\mathcal A_0
=
\begin{pmatrix}
0 & 0 & I_{3,0}\\
1 & 0 & I_{2,0}\\
0 & 1 & 0
\end{pmatrix}.
\label{eq:constant_companion}
\end{equation}

The corresponding differential equation therefore possesses constant
coefficients and admits a direct spectral interpretation.


The eigenvalues of the companion connection are determined by

\begin{equation}
\det(\lambda\mathbf 1-\mathcal A_0)
=
\lambda^3
-
I_{2,0}\lambda
-
I_{3,0}.
\label{eq:characteristic_poly}
\end{equation}

Denoting the three roots by

\begin{equation}
\lambda_1,\lambda_2,\lambda_3,
\label{eq:eigenvalues}
\end{equation}

one finds

\begin{equation}
\lambda_1+\lambda_2+\lambda_3=0,
\label{eq:sumroots}
\end{equation}

\begin{equation}
\lambda_1\lambda_2
+
\lambda_2\lambda_3
+
\lambda_3\lambda_1
=
-I_{2,0},
\label{eq:pairroots}
\end{equation}

\begin{equation}
\lambda_1\lambda_2\lambda_3
=
I_{3,0}.
\label{eq:prodroots}
\end{equation}


The constant Wilczynski invariants admit an immediate interpretation
in terms of Casimir data. Indeed,

\begin{equation}
C_2
=
\mathrm{Tr}(\mathcal A_0^2)
=
2I_{2,0},
\label{eq:C2}
\end{equation}

and

\begin{equation}
C_3
=
\mathrm{Tr}(\mathcal A_0^3)
=
3I_{3,0}.
\label{eq:C3}
\end{equation}

Thus the projective invariants obtained from the reduced BF
description are equivalent to the quadratic and cubic Casimir charges
of the companion connection.


The monodromy around the thermal circle is

\begin{equation}
M
=
\exp\!\left(
\beta\mathcal A_0
\right),
\label{eq:thermal_monodromy}
\end{equation}

whose eigenvalues are

\begin{equation}
m_i
=
e^{\beta\lambda_i},
\qquad
i=1,2,3.
\label{eq:monodromy_eigenvalues}
\end{equation}

Consequently, the global boundary data are completely encoded by the
conjugacy class of the monodromy matrix.


The constant Wilczynski invariants therefore determine the spectrum of
the companion connection through the characteristic equation
\eqref{eq:characteristic_poly}. Equivalently, they specify the
quadratic and cubic Casimir sectors of the reduced phase space. The
corresponding monodromy matrix \eqref{eq:thermal_monodromy} encodes
the global boundary data of the thermal configuration. In this way,
the pair $(I_{2,0},I_{3,0})$ provides a complete characterization of
the constant saddle sector. The thermodynamic consequences of these
monodromy data are investigated in the next section.
\section{Thermodynamics and Entropy}
\label{sec:entropy}

The following discussion is semiclassical and is intended to
illustrate the thermodynamic consequences of the monodromy spectrum
rather than provide a complete path-integral derivation.

The monodromy analysis of Section~\ref{sec:monodromy} provides a
direct bridge between the geometric data of the reduced
$\mathfrak{sl}(3,\mathbb R)$ BF theory and its thermodynamic
interpretation. In particular, the constant Wilczynski invariants
$(I_{2,0},I_{3,0})$ determine the spectrum of the companion
connection, while the associated monodromy eigenvalues characterize
the global thermal sector.


At the semiclassical level, the partition function is assumed to be
dominated by the constant saddle configuration,

\begin{equation}
Z(\beta,\mu_3)
\simeq
\exp
\!\left[
-I_{\rm cl}(\beta,\mu_3)
\right].
\label{eq:Z_semiclassical}
\end{equation}

Here $\beta$ denotes the inverse temperature and $\mu_3$ is the
chemical potential conjugate to the spin-3 charge.


For constant saddles, the classical action depends only on the
constant Casimir data. We therefore write the effective semiclassical
saddle ansatz

\begin{equation}
I_{\rm cl}
=
-\beta
\left(
\alpha I_{2,0}
+
g_3 I_{3,0}
\right),
\label{eq:Icl}
\end{equation}

where $\alpha$ and $g_3$ denote the effective spin-2 and spin-3
couplings, respectively. This expression should be understood as an
effective semiclassical saddle ansatz rather than as the result of a
full generalized Schwarzian path-integral evaluation. Its role is to
encode the constant spin-2 and spin-3 Casimir data of the reduced
phase space in the classical thermodynamic functional.

Consequently,

\begin{equation}
\log Z
=
\beta
\left(
\alpha I_{2,0}
+
g_3 I_{3,0}
\right).
\label{eq:logZ}
\end{equation}


The thermodynamic energy and spin-3 charge are defined by

\begin{equation}
E
=
-\frac{\partial}{\partial\beta}
\log Z,
\qquad
Q_3
=
\frac{\partial}{\partial\mu_3}
\log Z.
\label{eq:EQ3}
\end{equation}

These quantities are determined by the saddle values of the quadratic
and cubic Casimir invariants, which in the present formulation are
equivalently encoded by the constant Wilczynski data
$(I_{2,0},I_{3,0})$.


The entropy follows from the Legendre transform

\begin{equation}
S
=
\log Z
+
\beta E
-
\mu_3 Q_3 .
\label{eq:LegendreEntropy}
\end{equation}

The thermodynamic properties of the generalized Schwarzian theory are
therefore controlled by the monodromy spectrum of the companion
connection.


The physically relevant thermal sector corresponds to hyperbolic
monodromy. In this case the dominant eigenvalue is determined by the
characteristic equation

\begin{equation}
\lambda^3
-
I_{2,0}\lambda
-
I_{3,0}
=
0.
\label{eq:cubic_entropy}
\end{equation}

The corresponding monodromy eigenvalue is

\begin{equation}
m_{\max}
=
e^{\beta\lambda_{\max}}.
\label{eq:mmax}
\end{equation}

Since the largest eigenvalue governs the asymptotic growth of states,
the leading semiclassical entropy is expected to scale as

\begin{equation}
S_{\rm hyp}
\sim
\beta\lambda_{\max}.
\label{eq:S_lambda}
\end{equation}


Equation~\eqref{eq:cubic_entropy} is a depressed cubic equation. Its
discriminant is

\begin{equation}
\Delta_{\rm cub}
=
\frac{I_{3,0}^{\,2}}{4}
-
\frac{I_{2,0}^{\,3}}{27}.
\label{eq:cubic_discriminant}
\end{equation}

For the hyperbolic sector, the dominant real root is obtained from
Cardano's formula,

\begin{equation}
\lambda_{\max}
=
\left(
\frac{I_{3,0}}{2}
+
\sqrt{\Delta_{\rm cub}}
\right)^{1/3}
+
\left(
\frac{I_{3,0}}{2}
-
\sqrt{\Delta_{\rm cub}}
\right)^{1/3}.
\label{eq:lambda_max_cardano}
\end{equation}


Using Eq.~\eqref{eq:lambda_max_cardano}, the leading semiclassical
entropy may be written as

\begin{equation}
S_{\rm hyp}
\sim
\beta
\left[
\left(
\frac{I_{3,0}}{2}
+
\sqrt{\Delta_{\rm cub}}
\right)^{1/3}
+
\left(
\frac{I_{3,0}}{2}
-
\sqrt{\Delta_{\rm cub}}
\right)^{1/3}
\right].
\label{eq:entropy_final}
\end{equation}

This expression provides an explicit semiclassical relation between
the constant Wilczynski invariants and the entropy of the associated
hyperbolic thermal sector.


The ordinary Schwarzian result is recovered in the limit

\begin{equation}
I_{3,0}=0.
\label{eq:I3_zero}
\end{equation}

Equation~\eqref{eq:cubic_entropy} then reduces to

\begin{equation}
\lambda
\left(
\lambda^2-I_{2,0}
\right)
=
0,
\label{eq:sl2_limit_poly}
\end{equation}

and the dominant eigenvalue becomes

\begin{equation}
\lambda_{\max}
=
\sqrt{I_{2,0}}.
\label{eq:sl2_limit_lambda}
\end{equation}

Consequently,
\begin{equation}
S_{\rm hyp}
=
\beta\sqrt{I_{2,0}} .
\label{eq:sl2_limit_entropy}
\end{equation}
which reproduces the familiar Schwarzian scaling.


The thermodynamic sector is therefore controlled by the constant
Wilczynski invariants through their associated Casimir structures and
monodromy data. In this way, projective geometry, monodromy
invariants, and semiclassical entropy become directly linked within
the reduced $\mathfrak{sl}(3,\mathbb{R})$ BF theory.
\section{Discussion and Outlook}
\label{sec:discussion}

The main result of this work is a bulk-first derivation of
generalized Schwarzian dynamics from two-dimensional BF gravity.
Starting from flat BF connections and their associated boundary phase
spaces, we showed that Drinfeld--Sokolov reductions naturally lead to
projective differential equations whose invariant data are encoded by
the Wilczynski invariants. In the ordinary $\mathfrak{sl}(2,\mathbb R)$
case, this construction reproduces the familiar Schwarzian
derivative, while for $\mathfrak{sl}(3,\mathbb R)$ it yields a
generalized Schwarzian structure governed by the second and third
Wilczynski invariants.

A central conceptual aspect of the present construction is its
bulk-first character. Rather than introducing Schwarzian dynamics as
an independent boundary theory, we regard it as an emergent
description arising from reductions of the BF asymptotic phase space.
From this viewpoint, flat connections, residual gauge symmetries, and
Drinfeld--Sokolov reductions constitute the fundamental ingredients
of the construction, while Schwarzian and generalized Schwarzian
actions appear only after suitable boundary reductions have been
implemented.

The resulting framework also clarifies the geometric meaning of the
generalized Schwarzian variables. The quantities $I_2$ and $I_3$ are
not merely higher-derivative fields but intrinsic projective
invariants associated with curves in projective space. The reduced
boundary dynamics may therefore be interpreted geometrically in terms
of projective differential geometry, extending the familiar relation
between the Schwarzian derivative and projective structures
on the line to higher-rank settings \cite{OvsienkoTabachnikov,Wilczynski}.

Our analysis further establishes direct connections between generalized
Schwarzian dynamics, Casimir structures, monodromy data, and
semiclassical thermodynamics. In particular, the constant Wilczynski
invariants determine the quadratic and cubic Casimir sectors, which in
turn characterize the monodromy spectrum of the companion connection
and organize the corresponding thermal regime.

Another intriguing direction concerns the role of the dilaton sector.
While the connection sector gives rise to the Hill and Wilczynski
equations underlying the Schwarzian and generalized Schwarzian
actions, the corresponding dilaton multiplets appear to be governed
by higher-order stabilizer structures. Understanding this relation
more systematically may provide a geometric interpretation of
higher-spin dilaton sectors and their associated coadjoint orbits.

Several directions remain open. It would be particularly interesting
to derive the thermodynamic sector from a complete generalized
Schwarzian path integral and to determine the corresponding
one-loop corrections. Another natural extension is the analysis of
$\mathfrak{sl}(N,\mathbb R)$ BF theories 
\cite{Gutperle:2011kf,Perez:2013xi,deBoer:2013gz}, where higher Wilczynski
invariants are expected to generate an infinite hierarchy of
generalized Schwarzian structures. It would also be worthwhile to
investigate coadjoint-orbit quantization, higher-spin partition
functions, and the role of generalized Schwarzian dynamics in
low-dimensional holography. We hope that the bulk-first perspective
developed here will provide a useful framework for exploring these
questions.

\end{document}